\documentstyle[aps,prl,multicol,epsfig]{revtex}
\draft

\begin{document}

\title{Theory of Melting and the Optical Properties of 
Gold/DNA Nanocomposites}
\author{Sung Yong Park and D. Stroud}
\address{
Department of Physics,
The Ohio State University, Columbus, Ohio 43210}

\date{\today}

\maketitle

\begin{abstract}

We describe a simple model for the melting and optical
properties of a DNA/gold nanoparticle aggregate.
The optical properties at
fixed wavelength change dramatically at the melting transition, 
which is found to be higher and narrower in temperature
for larger particles, and much sharper than that of an isolated
DNA link.  All these features are in agreement with available
experiments.   The aggregate is modeled as a cluster of gold
nanoparticles on a periodic lattice connected by DNA bonds,
and the extinction coefficient is computed using the discrete
dipole approximation. 
Melting takes place as an increasing number of these bonds
break with increasing temperature.  The melting temperature
corresponds approximately to the bond percolation threshold.  

\end{abstract}


\begin{multicols}{2}
The optical properties of metallic nanoparticles have been 
investigated intensively over the last decades~\cite{bohren,kreibig}.
Recently, this investigation has expanded to include
so-called functional metallic nanoparticles\cite{sanchez}.
Among these, there is a particular interest in 
DNA-modified gold nanoparticles.  Such particles can form
complex aggregates which may melt at high temperatures.
The particles and their aggregates may have a variety of
applications, e.\ g., in biological detection, which 
may be possible by using 
the optical and electrical sensitivity of the 
aggregates~\cite{mirkin,elghanian,sjpark}.
Numerical model calculations of the optical properties of DNA 
modified gold nanoparticle aggregates show general agreement with 
experiments, including such features as the following:
(i) for isolated gold nanoparticles in suspension, there is a 
strong "surface plasmon" absorption in the visible; and (ii) 
this absorption maximum broadens and red-shifts when the 
cluster radius becomes 
comparable to the wavelength~\cite{lazar,storhoff1}. 
However, some important physical details have not been explained.  
For example, in the gold/DNA nanoparticle system, 
the melting transition of a gold nanoparticle aggregate has a 
much narrower temperature width and 
occurs at a high temperature than that of a single DNA 
link~\cite{elghanian,drukker}. Also, the dependence of aggregate 
melting temperature on particle size has not yet been 
explained~\cite{kiang}.

In this Letter, we  model this novel ``melting'' transition of
a gold/DNA nanoparticle aggregate. 
Our model accounts for most experimentally observed features, 
including (i) the small temperature width of the aggregate
melting transition, compared to that of a single DNA link; 
(ii) the particle-size-dependence of the melting temperature $T_m$;
and (iii) the temperature-dependence of the optical extinction 
coefficient $C_{ext}(\lambda,T)$ at wavelength $\lambda$ and 
temperature $T$. 

We first describe our model for the low-temperature
morphology of the aggregate.
In the simplest version of our model, the low-T cluster is taken 
simply as a collection of identical gold nanoparticles 
(each of radius $a$) 
which fill the sites of a simple cubic lattice of lattice
constant $d$ ($d > 2a$).  The cluster is assumed to be 
a cube of edge $L$,
containing $N_{\rm par} = (L/d)^3$ gold nanoparticles.  We 
have also investigated the melting and the optical properties 
assuming that the low-T cluster is a fractal aggregate.

To describe the melting of this aggregate, we assume that each nanoparticle carries exactly $N_s$ single 
DNA strands.  (This choice still retains the essential features of
the melting.)  The cluster exists at low $T$ because a chemical 
reaction converts two single strands $S$ into a double strand $D$.
In this Letter,  we adopt a simplified two-state model
for this reaction\cite{bloomfield,note1}, described by the relation
\begin{equation}
S + S \rightleftharpoons D.
\label{eq:react}
\end{equation}
For short DNA strands (12-14 base pairs), this model describes 
melting well\cite{werntges}.
Each gold nanoparticle has $z$ nearest neighbors 
($z = 6$ for a simple cubic lattice).  Hence, there
are $N_{\rm par}z/2$ bonds joining adjacent nanoparticles.
We assume that a (temperature-dependent) fraction $p(T)$ of
the single DNA strands  
form double strands by the reaction (\ref{eq:react}).  In order
to calculate the fraction $p_{\rm eff}(T)$ of {\em bonds} which
contain at least one double strand, we adopt the following 
model.  First, we assume that exactly $N_s/z$
of the single strands on a given nanoparticle are available
to bond with any {\em one} of its z nearest neighbors.  
Thus, we assume that the maximum number of links that could be 
formed
between any two particles is $N_s/z$. The probability that
{\em no} link is formed is then taken to be
\begin{equation}
1 - p_{\rm eff}(T) = [1 - p(T)]^{N_s/z}.
\end{equation}

The criterion for the melting temperature $T_c$ is that 
$p_{\rm eff}(T_c) = p_c$,
$p_c$ being the bond percolation threshold\cite{stauffer} for the 
lattice considered, 
at which an infinite connected path of double DNA strands first 
forms.
(For example, $p_c \sim 0.25$ on a very large simple cubic
lattice.)  Thus, at percolation,
\begin{equation}
[1 - p(T_c)]^{N_s/z} = 1 - p_c.
\label{eq:melt}
\end{equation}
This equation implicitly determines $T_c$ in terms of $z$, $p_c$, 
and $N_s$\cite{note2}.  

$p(T)$ itself is determined by a chemical equilibrium between the 
single-strand and double-strand DNA molecules, 
which are attached to the gold nanoparticles.  
The chemical equilibrium condition corresponding to 
(\ref{eq:react}) is
\begin{equation}
\frac{[1-p(T)]^2}{p(T)} = \frac{K(T)}{C_T} \equiv K^\prime(T),
\label{eq:equil}
\end{equation}
where $K(T)$ is a suitable chemical equilibrium constant, 
and $C_T$ is the molar concentration of single DNA strands in
the sample (in our model all are attached to the 
nanoparticles).
Since $0 < p(T) < 1$,
the physical solution to eq.\ (\ref{eq:equil}) is
$p(T) = 1 + \frac{1}{2}\left(K^\prime - \sqrt{K^{\prime,2} + 
4K^\prime}\right)$.

Note that, according to eq.\ (\ref{eq:melt}), $p(T_c)$ decreases 
with increasing $N_s$ and, hence, with increasing particle radius. 
Since $p(T)$ decreases monotonically with $T$, 
$T_c$ should thus be an {\em increasing} function of $a$, 
as reported in experiments\cite{kiang}.  

\begin{figure}
\epsfxsize=8cm \epsfysize=6cm \epsfbox{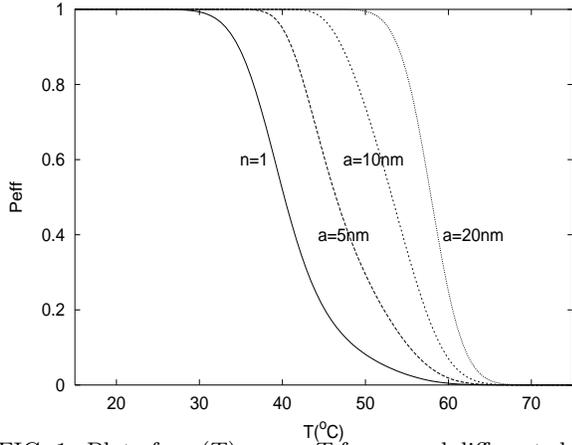} 
\caption{Plot of $p_{\rm eff}(T)$ versus $T$ for several different
choices of particle radius $a$, as indicated.
Also plotted is p(T), the probability that a given DNA strand 
is part of a double strand at $T$.}
\label{fig:1}
\end{figure}

In Fig.\ \ref{fig:1}, we plot p$_{\rm eff}$(T) for several $a$.  We
assume $N_s \propto a^2$, set $z = 6$ and use the 
experimental result that $N_s = 160$ 
when $a = 8$ nm~\cite{demers}.  
We have also assumed 
the simple van't Hoff behavior $K(T) = 
\exp[-\Delta G/k_BT]$, with a
Gibbs free energy of formation
$\Delta G(T) = c_1(T - T_M) + c_3(T - T_M)^3$, choosing
the values of $c_1$, $c_3$, and $T_M$ to be consistent
with experiments on these DNA molecules.

\begin{figure}
\epsfxsize=8cm \epsfysize=6cm \epsfbox{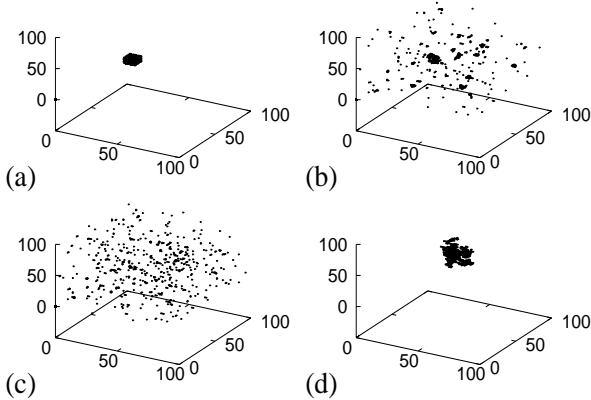} 
\caption{Schematic of the melting of a gold-DNA cluster, for
two different models discussed in the present paper.  In
the first model, (a) at low $T$ the cluster is described
as an $L \times L \times L$ simple cubic aggregate of lattice
constant $d$ [here $L/d = 10$].  
(b) As the temperature $T$ increases, some of the bonds break.  
The fraction of bonds present is $p_{\rm eff}(T)$.  In (b),
$p_{\rm eff}=0.5 > p_c(L)$, 
where $p_c$ is the weakly $L$-dependent percolation threshold.
A percolation cluster of linear dimension $\sim L$ still exists.
At a still higher $T$ [shown in (c)], $p_{\rm eff}(T) = 0.2 < p_c(L)$,
and only small clusters remain.  
In our calculation, the clusters are positioned and oriented at 
random in the bounding box, taken as a cube of edge 100$d$.
(d) Alternate model for low-$T$ sample ($p_{\rm eff}(T) = 1$):
fractal cluster formed by
reaction-limited cluster-cluster aggregation [RLCA],  
with fractal dimension $d_f \sim 2.1$.}
\label{fig:2}
\end{figure}

To calculate the $T$-dependent optical 
properties, we have considered the melting of 
two slightly different
low-temperature aggregates.  In the first, we 
assume that, at low $T$, the aggregate consists of a simple 
cubic collection of $N_{\rm par} = (L/d)^3$ gold nanoparticles on 
a simple cubic lattice, 
as described above, with all bonds occupied by DNA double-strand 
links.  To generate a specific sample with a given p$_{\rm eff}$(T), 
we randomly remove links 
with probability $1 - p_{\rm eff}(T)$, then identify the separate 
clusters, using a simple computer algorithm\cite{stauffer}.  
We neglect gravitational forces, which may be important in
some experimental circumstances. 
If the aggregate consists of two or more clusters, we simply 
place these aggregates in random positions and 
orientations within a larger bounding box (usually of edge 100$d$),
 taking care that the individual clusters do not overlap.   The 
resulting geometry is shown schematically
in Fig.\ \ref{fig:2}.  
We have also carried out the same procedure 
to simulate the melting of a sample formed by reaction-limited 
cluster-cluster aggregation 
(RLCA)~\cite{brown}.  A typical RLCA cluster is shown
in Fig.\ \ref{fig:2} (d), and represents a
possible fractal aggregate which might be produced by 
certain random growth processes at low $T$\cite{sypark}.  

We calculate the optical properties of this sample using
the Discrete Dipole Approximation (DDA)\cite{pp,draine}.  
The sample is modeled as a collection of separate
aggregates whose extinction coefficients are computed individually,
then added.  Each aggregate consists of many identical 
nanoparticles, which have complex frequency-dependent 
dielectric constant $\epsilon(\omega)$, and polarizability
$\alpha(\omega)$ related to $\epsilon(\omega)$
by the Clausius-Mossotti equation\cite{purcell1}
$\epsilon(\omega) = 1 + 4\pi n\alpha/[1 - (4\pi n/3)\alpha]$,
where $n =(d/L)^3$. 
The resulting expressions for the induced dipole moment 
${\bf p}_i$ of the $i^{th}$ sphere, and the corresponding expression
for the extinction coefficient $C_{ext}(k)$ at wave number $k$,
are given in Ref.\ \cite{lazar}.

In our case, each cluster consists of a number of DNA-linked 
individual gold nanoparticles.
In our calculations, we do not include
the optical properties of the DNA molecules, since
these absorb primarily in the ultraviolet~\cite{storhoff1}.  
We use tabulated values of the gold complex
index of refraction\cite{lynch,johnson}, then  
calculate $C_{ext}$ for each cluster
using the DDA.  To improve the statistics,
we average $C_{ext}$ for each cluster over possible orientations.
We then sum the averaged extinction coefficients of all the
individual clusters to get the total extinction coefficient 
of the suspension.
This method is justified when the suspension is dilute.

\begin{figure}
\epsfxsize=8cm \epsfysize=6cm \epsfbox{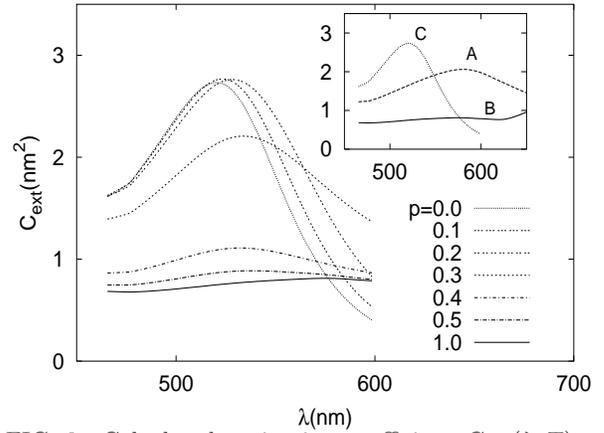} 
\caption{Calculated extinction coefficient $C_{ext}(\lambda, T)$
as a function of wavelength $\lambda$, plotted for several
values of $p_{\rm eff}$, particle radius 20 nm, as indicated in
the legend.  $p_{\rm eff} = 1$ corresponds  to 1000 gold nanoparticles on 
a simple cubic lattice of edge $L = 480$ nm.  The sample is that shown 
schematically in Fig.\ \ref{fig:2} (a) - (c).  
Inset: $C_{ext}(\lambda, T)$ for the
RLCA aggregate at $p_{\rm eff} = 1$ (curve A), as well as for simple 
cubic aggregate at $p_{\rm eff} = 1$ (curve B), 
and for individual gold nanoparticles
$p_{\rm eff} = 0$.(curve C).}
\label{fig:3}
\end{figure}

In Fig.\ \ref{fig:3}, we show $C_{ext}(\lambda, T)$ versus 
$\lambda$ for the gold-DNA 
cluster at several values of p$_{\rm eff}$(T), assuming 
$a = 20$ nm.
p$_{\rm eff}$ = 0 represents a dispersion of $N_{\rm par} = (L/d)^3$ 
individual nanoparticles, 
while p$_{\rm eff}$ = 1 represents a simple cubic lattice of connected 
nanoparticles. 
The calculated $C_{ext}(\lambda, T)$ at each $\lambda$ changes
strikingly for 
p$_{\rm eff}$(T) $\sim$ p$_c$. For small p$_{\rm eff}$, 
there is a clear extinction peak near 
$520$ nm.  This peak corresponds to the 
wavelength of the surface plasmon resonance 
(SPR) in individual Au nanoparticles.  
As p$_{\rm eff}$ increase, this peak first 
red-shifts, then greatly broadens, as the aggregate melts.
The calculated peak is, however, shifted much more and is much 
broader than experiment.  However, as shown in inset, 
if we assume that the low-$T$ aggregate is an RLCA fractal,
the calculated peak shift is consistent with 
experiment~\cite{sypark}.

\begin{figure} 
\epsfxsize=8cm \epsfysize=6cm \epsfbox{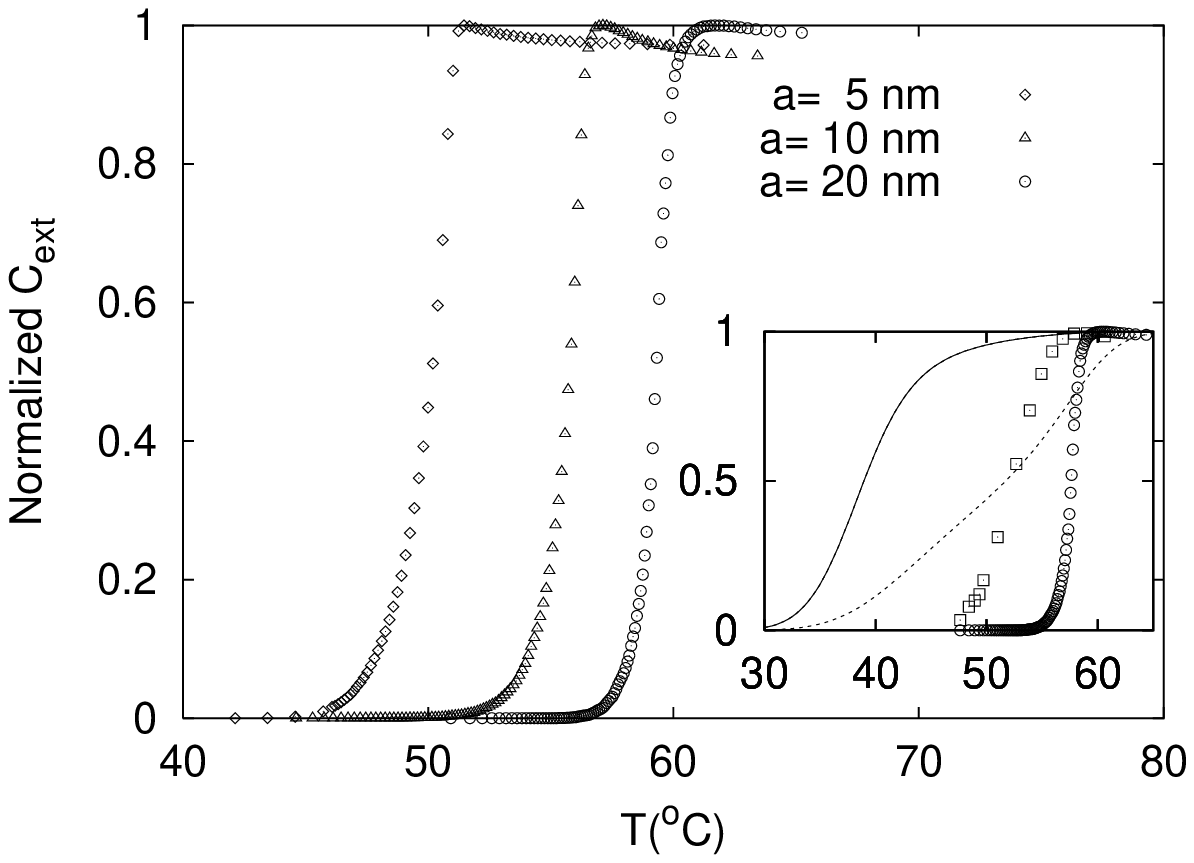} 
\caption{Normalized extinction coefficient $C_{ext}(\lambda, T)$
for $\lambda = 520$ nm, plotted versus $T$ 
for several particle radii $a$, assuming
that the low-$T$ aggregate is an $N = 1000$ simple cubic
cluster, as shown in Fig.\ \ref{fig:2}. 
Inset: Normalized extinction coefficient $C_{ext}(\lambda, T)$,
versus $T$ for $\lambda = 520$ nm, plotted for a 1000-particle
gold/DNA aggregate assuming that the low-temperature
sample is a simple cubic cluster (open circles) or
an RLCA cluster (open squares). The solid curve is a plot
of $1 - p(T)$ for a single DNA duplex with the same
concentration $C_T$ as the above two curves.
The dotted curve represents $1 - p(T)$ for a 
single DNA duplex but with a much higher $C_T$ than for
the other curves of the inset.  The plots of $1 - p(T)$ for a
single link closely resemble $C_{ext}(\lambda, T)$ as 
measured at $\lambda = 260$ 
nm for a single duplex[11].}

\label{fig:4}
\end{figure}

In Fig.\ \ref{fig:4}, we show $C_{ext}(\lambda, T)$ at {\em fixed}
$\lambda = 520$ nm (close to the isolated-particle SPR),
versus $T$, for several particle sizes, 
assuming that the low-T aggregate is a 
simple cubic cluster with $N = 1000$ particles, 
as shown in Fig.\ \ref{fig:2}.
For each $a$, the extinction increases sharply at a characteristic
$T$, corresponding to the melting of the aggregate for that
$a$; at this $T$, the absorption due to the SPR increases
sharply.   

In the inset of Fig.\ \ref{fig:4}, we compare 
$C_{ext}(\lambda, T)$ for a regular and an 
RLCA cluster of particles of $20$ nm radius at 
$\lambda$ = 520 nm, both for $N = 1000$ 
particles.  Although the RLCA cluster has a 
slightly broader melting transition, as 
manifested in $C_{ext}(\lambda, T)$, than does the regular lattice,
both sets of data show a much sharper melting transition than
that of a single DNA link.  Also, although our normalized
$C_{ext}(\lambda, T)$ is calculated for the aggregates
at 520 nm, we expect similar behavior at $260$ nm. 
(We have not carried out calculations at this $\lambda$ mainly 
because we have not included the DNA absorption properties.)  
In any case, the experimental melting curves at 260 and 520 nm 
are very similar\cite{kiang}.

Our calculated extinction coefficients are strikingly similar
to recent experimental results\cite{elghanian,storhoff1,kiang}.
In particular, both experiment and calculation give a sharp increase 
in $C_{ext}(\lambda, T)$ at fixed $\lambda$, as $T$ increases 
past a critical 
temperature, which we interpret as the melting temperature $T_m$.
We also find, in agreement with 
experiment\cite{elghanian,storhoff1}, that 
melting occurs over a much narrower range of $T$ in the
aggregate than for a single bond, and that the melting occurs
at higher $T$ for larger particles\cite{kiang}.  The crucial
point is that $p_{\rm eff}(T)$ is a much sharper function of $T$
than is $p(T)$, and this feature would not be affected by slight 
changes
in the model (such as considering a body-centered-cubic
rather than a simple cubic cluster).
In the present model, the melting transition is in 
the universality class of bond percolation.      
This universality class might change because of the constant 
dissolving and reforming of
the bonds of the aggregate (a possibility omitted from our model).  
This possibility should be studied further 
theoretically and experimentally.
However, the optical properties 
may well be robust, in the sense 
that $C_{ext}(T)$ would be little affected even if this feature 
were included, since the key ingredient 
is the multiple links per bond as described above.

In summary, we have developed a simple model for the melting of aggregates of
gold nanoparticles and DNA, and have calculated the {\em T-dependent} optical
properties of these melting aggregates, using the DDA.  We find that, 
at fixed $\lambda$, melting is accompanied by dramatic changes
in the extinction coefficient $C_{ext}(\lambda)$.  
These calculated changes occur over a {\em much narrower}
temperature range than that over which interparticle links 
themselves melt, the temperature width is smaller for 
particles with larger radius, and the melting occurs at
{\em higher temperatures} for the {\em larger} 
nanoparticles than for the smaller ones.  
All these effects are in good agreement with 
experiment\cite{elghanian,storhoff1,kiang}.
It would be of great value if the predicted $T$-dependent structure
of the clusters could be probed experimentally, e. g. by
static light scattering, as has been done in other
contexts\cite{lin}.  Finally, we mention that the present model
can be applied to other similar nanocomposites\cite{mann}.

This work has been supported by grant NSF DMR01-04987, by the
U. S./Israel Binational Science Foundation, and by
an Ohio State University Postdoctoral Fellowship awarded to
S.\ Y.\ Park.   We thank D. J. Bergman, C.-H. Kiang, A.\ A.\ Lazarides, and D.\
R.\ Nelson for valuable conversations.

\end{multicols}

\begin{references}
\bibitem{note1} In actuality, there is a linker molecule which 
emerges from solution to connect two DNA single strands, each
on a different nanoparticle. Even in this case, however,
the melting condition can be reduced to that used here, without
loss of accuracy, as will be shown elsewhere.  Also, 
there are two chemically distinct DNA single strands 
(denoted A and B); 
each gold particle has either all A or all B single strands 
attached, and linker
molecule can only attach A and B single strands.
The existence of two species can be disregarded by symmetry when the 
the concentrations of the two species in solution are the same; 
experiments have been carried out only under this condition.
\bibitem{note2} We implicitly assume that the 
specific bonds which are occupied at 
temperature $T$ are time-independent.  In reality, these links 
slowly break up and reform in time.   The effects of this
assumption are discussed further below.
\end{references}
\end{document}